\documentclass[prd,twocolumn,nofootinbib,aps,superscriptaddress,tightenlines]{revtex4}

\usepackage{graphicx}

\def\nn{\nonumber \\ }
\def\rep#1#2#3{(\mathbf{#1},\mathbf{#2})_{#3}}

\def\vev#1{\left\langle #1 \right\rangle}
\def\tr{\text{Tr}\,}

\begin{document}

\preprint{ \vbox{\hbox{UCSD/PTH 06-07} \hbox{CALT-68-2601}  } 
}

\title{Flavor Changing Neutral Currents, an Extended Scalar Sector, and the Higgs Production Rate at the LHC}

\author{Aneesh V. Manohar}

\affiliation{Department of Physics, University of California at San Diego, La Jolla, CA 92093}

\author{Mark B. Wise}

\affiliation{California Institute of Technology, 452-48, Pasadena, CA 91125 }

\begin{abstract}
We study extensions of the standard model with additional colored scalar fields which can couple directly to quarks. Natural suppression of flavor changing neutral currents implies minimal flavor violation, and fixes the scalars to transform as $\rep82{1/2}$ under the $SU(3) \times SU(2) \times U(1)$ gauge symmetry. We explore the phenomenology of the standard model with one additional $\rep82{1/2}$ scalar, and discuss how this extension can modify flavor physics and the Higgs boson production rate at the LHC. Custodial $SU(2)$ symmetry can be implemented for the octet scalars since they transform as a real color representation. Additional weak scale degrees of freedom needed for gauge unification are discussed.
\end{abstract}

\date{\today}

\maketitle

%
\section{Introduction}
\label{sec:intro} 

The minimal standard model for strong weak and electromagnetic interactions has been extremely successful. However, the recent measurement of neutrino masses requires some modifications to the minimal theory, the simplest being the addition of nonrenormalizable dimension five operators which generate neutrino masses and mixings. The smallness of the neutrino masses, and the fact that in the minimal standard model the couplings almost unify~\cite{unify}, suggests that unification of the strong weak and electromagnetic interactions occurs at some high scale, and neutrino mass terms are generated by the seesaw mechanism~\cite{seesaw}.

Most of the extensions of the standard model with new physics at the TeV scale have been motivated by the hierarchy puzzle, i.e., why is the weak scale so small compared with the Planck or unification scales. However, the measured value of the cosmological constant suggests that a fine tuning that is qualitatively similar to that needed to achieve the smallness of the weak scale is needed for the cosmological constant. Perhaps we are not looking at this issue correctly.

If one does not adopt the hierarchy puzzle as the criteria for motivating extensions of the standard model then one can take a more general point of view. Certainly the hypothesis that the scalar sector contains  a single scalar doublet has not been tested experimentally. Adding additional scalars that can couple to the quarks typically results in unacceptably large flavor changing neutral currents.\footnote{We use the phrase ``flavor changing neutral currents'' even though scalars couple to quark bilinears that are scalars and not four vectors} Glashow and Weinberg~\cite{Glashow} studied extensions with many scalar doublets, and found that if one scalar coupled to the up type quarks and one scalar to the down type quarks, then large tree level flavor changing neutral currents are avoided. They noted that this can be enforced by discrete symmetries. 

In this paper, we study the most general scalar structure of the standard model, which naturally maintains the smallness of flavor changing neutral currents.
In the minimal standard model, the Yukawa interactions of the Higgs doublet with the quarks are the only interactions that violate the $SU(3)_U \times SU(3)_D \times SU(3)_Q$ flavor symmetry. Here $SU(3)_U$ acts on the right-handed up type quarks, $SU(3)_D$ on the right handed down type quarks, and $SU(3)_Q$ on the left handed quark doublets. By studying the breaking of this flavor symmetry, we show in Sec.~\ref{sec:fcnc} that the smallness of flavor changing neutral currents implies the principle of minimal flavor violation (MFV)~\cite{Georgi}, and restricts the scalars that can Yukawa couple to the quarks to either: (1) have the same gauge quantum numbers as the Higgs doublet, or (2) transform under $SU(3) \times SU(2) \times U(1)$ as $\rep82{1/2}$. The phenomenology of the minimal standard model with an additional $\rep82{1/2}$ scalar is studied in the remainder of this paper. This model contains additional sources of $CP$ nonconservation beyond the standard model CKM phase. The new sources of $CP$ nonconservation give rise to a contribution to the electric dipole moment of the neutron that can be near its present experimental limit. The phenomenology of color octet scalars (in the context of Pati-Salam unification) has been studied in Ref.~\cite{Popov}.

In a previous paper~\cite{Manohar}, we showed that the production rate for the Higgs particle at the LHC is very sensitive to new physics. In this paper we show that for a range of parameters, the color octet scalars we have added can have a significant impact on the rate for Higgs production through gluon fusion. Even though there are no flavor changing neutral currents from tree level color octet scalar exchange, the charged octet scalars induce flavor changing effects through loop graphs. These small deviations in the standard model predictions may be of interest in future high precision tests of the flavor structure of the standard model.

There are two perspectives one can have on the model we are studying. One is to treat the scalars as fundamental up to some very high scale. In this case, it is desirable to have unification of the gauge couplings. We show that if one adds three real scalars with $SU(3) \times SU(2) \times U(1)$ quantum numbers $\rep130$ then acceptable unification is achieved with a unification scale that is large enough to suppress proton decay. On the other hand it is possible that the scalars observed are only the low energy remnants of some more complicated dynamics for the scalar sector that becomes manifest at an energy scale much below the GUT or Planck scales.

The next section outlines constraints on the $SU(3) \times SU(2) \times U(1)$ quantum numbers of additional scalars that couple to quarks and naturally have no tree level flavor changing neutral currents. The explicit model we consider is given in Section \ref{sec:model}. Section \ref{sec:matching} discusses the effects of the octet scalars on precision electroweak physics, and on the Higgs production cross-section. Flavor physics and the electric dipole moment of the neutron are discussed in sections~\ref{sec:flavor} and~\ref{sec:dipole}. In section~\ref{sec:production}, the production cross section of the color octet scalars at the LHC is computed at leading order. Some possible decays of these states are briefly discussed. The impact of the new scalars on unification of the coupling constants is studied in Section \ref{sec:unify}. Concluding remarks are given in section \ref{sec:conclude}.

\section{Flavor Changing Neutral Currents and Minimal Flavor Violation}
\label{sec:fcnc}

We will consider a generalization of the standard model, where we include, in addition to the usual Higgs doublet $H$, all possible scalar representations which can couple to quarks. We require that the theory naturally have suppressed flavor changing neutral currents. This requirement, that flavor conservation by neutral currents ``follows from the group structure and representation content of the theory, and does not depend on the values taken by the parameters of the theory'' was formulated by Glashow and Weinberg~\cite{Glashow}. We will see that it greatly limits the possible scalar content of the theory and allowed scalar couplings.

The standard model quark Yukawa couplings are 
\begin{eqnarray}
L &=& -g_{ij}^U \bar u_{Ri}  Q_j H- g_{ij}^D \bar d_{Ri}  Q_j H^\dagger + \text{h.c.},
\label{yuk}
\end{eqnarray}
where $i$ and $j$ are flavor indices, and gauge indices have been omitted. The Yukawa couplings generate the mass matrices
\begin{eqnarray}
M_{ij}^U  = g_{ij}^U \vev{H},\qquad  M_{ij}^D = g_{ij}^D \vev{H}^\dagger,
\label{mass}
\end{eqnarray}
for the charge $2/3$ and $-1/3$ quarks  when the Higgs field gets a vacuum expectation value $\vev{H}=v/\sqrt{2}$. If there are no tree-level flavor changing neutral currents in the Higgs sector, one requires that the Yukawa couplings for the neutral Higgs boson be diagonal in the basis where the quark mass terms are diagonal~\cite{Glashow}. This is automatically satisfied by Eq.~(\ref{yuk}), since the Yukawa couplings and mass matrices are proportional to each other. The standard model Lagrangian has an $SU(3)_U \times SU(3)_D \times SU(3)_Q$ flavor symmetry under which the Yukawa couplings can be considered as sources which transform as $g^U \sim (\mathbf{3}_U,\mathbf{\bar 3}_Q)$, $g^D \sim (\mathbf{3}_D,\mathbf{\bar 3}_Q)$ so that Eq.~(\ref{yuk}) preserves the flavor symmetry. The flavor symmetry is explicitly broken by setting $g^{U,D}$ equal to fixed matrices.

Now consider the most general set of scalars that can Yukawa couple to standard model quarks through either quark bilinears or quark-lepton bilinears. The possible scalar representations are $\rep12{1/2}$, $\rep82{1/2}$, $\rep63{1/3}$, $\rep61{4/3,1/3,-2/3}$, $\rep33{-1/3}$, $\rep31{2/3,-1/3,-4/3}$, and their complex conjugates, and there may be multiple scalars of each type. Natural neutral current flavor conservation requires that the neutral scalars in all these multiplets automatically have diagonal Yukawa couplings in the basis in which the quark mass matrix is diagonal. 

The case of multiple Higgs doublets in the $\rep12{1/2}$ representation was studied by Glashow and Weinberg~\cite{Glashow}, and we summarize their argument here: One has a set of fields $H_\alpha$ and Yukawa matrices $g^{U,D}_\alpha$, and the fermion mass matrices are
\begin{eqnarray}
M^U  =\sum_\alpha g_\alpha^U \vev{H_{\alpha}},\qquad  M^D = \sum_\beta g_\beta^D \vev{H_{\beta}}^\dagger.
\label{mass2}
\end{eqnarray}
We require that $g^{U,D}_\alpha$ be diagonal in the same basis where $M^{U,D}$ are diagonal. A natural way this is possible is for there to be only one term in the sum for $M^U$ and only one term in the sum for $M^D$, i.e.\ only one scalar doublet couples to the $u$-type quarks, and one scalar doublet couples to the $d$-type quarks. The non-zero terms in Eq.~(\ref{mass2}) can have $\alpha=\beta$, or $\alpha\not=\beta$. This is the Glashow-Weinberg criterion for Higgs couplings~\cite{Glashow}.

We now analyze the Glashow-Weinberg criterion in a group-theoretic way. We will see that it leads to a more general solution than the one in Ref.~\cite{Glashow}, and also implies the MFV principle ~\cite{Georgi}. The group-theoretic analysis then generalizes trivially to the case of colored scalars. 

$g^U_\alpha$ and $M^U$ transform as $(\mathbf{3}_U,\mathbf{\bar 3}_Q)$ under the flavor symmetry, and $g^D_\alpha$ and $M^D$ transform as $(\mathbf{3}_D,\mathbf{\bar 3}_Q)$. We require that $g^U_\alpha$ are simultaneously diagonal in basis where $M^U$ is diagonal, i.e.\ that $g^U_\alpha$ and $M^U$ all break $SU(3)_U \times SU(3)_Q$ to the same diagonal $U(1)^2$ subgroup. The only natural way\footnote{See Appendix~\ref{app:natural} for some comments on naturalness.} this is possible is for there to be one independent symmetry breaking matrix $G^U$ transforming as $(\mathbf{3}_U,\mathbf{\bar 3}_Q)$, with
$g^U_\alpha \propto G^U$, $M^U \propto G_U$. Similarly, there is one symmetry breaking matrix $G^D$ transforming as $(\mathbf{3}_D,\mathbf{\bar 3}_Q)$, with
$g^D_\alpha \propto G^D$, $M^D \propto G_D$. In summary, all the flavor symmetry breaking terms must be proportional to the matrices $G^{U,D}$. This is sometimes referred to as MFV~\cite{Georgi}; we see that it is a consequence of imposing natural flavor conservation of neutral currents. It allows for Eq.~(\ref{mass2}) with several terms in each sum, provided all the Yukawa matrices are proportional to $G^U$ and $G^D$.

In our case, one has Yukawa coupling matrices for all the allowed scalar representations which can couple to quarks. Natural neutral current flavor conservation requires that all these Yukawa couplings must be obtained from the basic invariants $G^{U,D}$, so that they are flavor diagonal in the basis where $G^{U,D}$ (and hence $M^{U,D}$) are flavor diagonal. This requirement eliminates most of the allowed scalar representations. For example $\phi=\rep63{1/3}$  couples to $Q_iQ_j$, and has a Yukawa coupling which transforms as $\mathbf{\bar 6}_Q$ under flavor, since the symmetric product $Q_i Q_j$ transforms as $\mathbf{6}_Q$ under flavor.  It is not possible to make a $\mathbf{\bar 6}_Q$ from any product of $G^U \sim (\mathbf{3}_U,\mathbf{\bar 3}_Q)$ and $G_D \sim (\mathbf{3}_D,\mathbf{\bar 3}_Q)$. A simple way to see this is to look at triality under the diagonal $SU(3)$ group. $G^{U,D}$ have triality 0, and $\mathbf{\bar 6}_Q$ has triality one. The only representations which have Yukawa couplings to quarks with natural flavor conservation are the $\rep12{1/2}$ and $\rep82{1/2}$, and the allowed couplings automatically satisfy the MFV constraint. The $\rep82{1/2}$ Yukawa couplings have the same form as the $\rep12{1/2}$ couplings in Eq.~(\ref{yuk}) except for the color structure and an overall normalization,
\begin{eqnarray}
L &=& - \eta_U g_{ij}^U \bar u_{Ri}T^A  Q_j S^A- \eta_D g_{ij}^D  \bar d_{Ri} T^A Q_j S^{A\,\dagger}+\text{h.c.}~,\nn
\label{yuk8}
\end{eqnarray}
where $\eta_{U,D}$ are (complex) constants.

The above analysis does not readily generalize to leptons because of incomplete information on the neutrino sector of the theory. There are strong experimental constraints on flavor changing neutral currents in the charged lepton sector, from limits on processes such as $K \to \mu e$, $\mu \to e \gamma$ and $\mu \to 3 e$, but no comparable limits on flavor changing neutral currents in the neutrino sector. It is also not known whether there are only light doublet neutrinos with Majorana masses, or whether there also exist light singlet neutrinos. The possible scalars which can couple to lepton bilinears (including possible singlet neutrinos) have the quantum numbers
$\rep12{1/2}$, $\rep13{1}$, and $\rep11{2,1,0}$.

\section{The Model}
\label{sec:model}

The quantum numbers of scalars that can couple to quarks, and hence influence precision flavor physics, are very constrained by the principle of MFV. From the point of view of LHC phenomenology, the case of additional color octets is particularly interesting because virtual color octets that couple to the standard model Higgs doublet through the scalar potential can have a significant impact on the rate for Higgs production through gluon fusion. Motivated by this we study a simple model that adds to the minimal standard model a $\rep82{1/2}$ scalar multiplet 
\begin{eqnarray}
S^A &=& \left( \begin{array}{cc} {S^{+}}^A \\ {S^{0}}^A \end{array}\right),
\end{eqnarray}
where $A=1,\ldots,8$ is an adjoint color index. The usual $\rep12{1/2}$ Higgs doublet is denoted by $H$.

The most general renormalizable scalar potential is,

\begin{eqnarray}
V &=& \frac{\lambda}{4}\left(H^{\dagger i} H_i-\frac {v^2}{2}\right)^2 + 2 m_S^2 \tr S^{\dagger i} S_i \nn
&& +
\lambda_1 H^{\dagger i} H_i \tr S^{\dagger j} S_j 
+ \lambda_2 H^{\dagger i} H_j \tr S^{\dagger j} S_i\nn
&& + \Bigl[ \lambda_3  H^{\dagger i} H^{\dagger j} \tr S_i S_j  + \lambda_4 H^{\dagger i} \tr S^{\dagger j} S_j  S_i \nn
&& + \lambda_5 H^{\dagger i} \tr S^{\dagger j} S_i  S_j + \text{h.c.}\Bigr]\nn
&&+ \lambda_6 \tr S^{\dagger i} S_i S^{\dagger j} S_j 
+ \lambda_7 \tr S^{\dagger i} S_j S^{\dagger j} S_i \nn
&& + \lambda_8 \tr S^{\dagger i} S_i  \tr S^{\dagger j} S_j 
+ \lambda_9 \tr S^{\dagger i} S_j  \tr S^{\dagger j} S_i \nn
&&+\lambda_{10} \tr  S_i   S_j  \tr S^{\dagger i} S^{\dagger j}
+ \lambda_{11} \tr  S_i  S_j  S^{\dagger j} S^{\dagger i} .\nn 
\label{pot}
\end{eqnarray}
We have explicitly displayed the $SU(2)$ indices on the Higgs doublet and on the color octet scalars. Traces are over color indices and the notation, $S=S^AT^A$ is used. 
$\lambda_3$ has been made real by a phase rotation of the $S$ fields. With this phase convention, the phases of $\eta_{U,D}$, and $\lambda_{4,5}$ represent additional sources of CP violation beyond those in the minimal standard model.

The custodial $SU(2)$ symmetry of the standard model can be extended to this theory, since $S^A$ is an $SU(2)$ doublet and a \emph{real} representation under color. One can construct\footnote{This construction would not be possible if the scalars transformed under a complex color representation since the different components of $\mathcal{S}$ would transform differently under color.}
\begin{eqnarray}
\mathcal{S}^A&=&\left( \begin{array}{cc} {S^{0*}}^A & {S^{+}}^A \\
-{S^{-}}^A & {S^{0}}^A
\end{array}
\right)
\end{eqnarray}
in analogy with the Higgs doublet
\begin{eqnarray}
\mathcal{H}&=&\left( \begin{array}{cc} {H^{0*}} & {H^{+}} \\
-{H^{-}} & {H^{0}}
\end{array}
\right)
\end{eqnarray}
which transform under custodial $SU(2)$ as
\begin{eqnarray}
\mathcal{H} &\to& U \mathcal{H} U^\dagger, \nn
\mathcal{S}^A &\to& U \mathcal{S}^A U^\dagger.
\end{eqnarray}
This symmetry is left unbroken by the Higgs vacuum expectation value
\begin{eqnarray}
\vev{\mathcal{H}} &=& \left( \begin{array}{cc} \frac{v}{\sqrt{2}} & 0 \\
0 & \frac{v}{\sqrt{2}}
\end{array}\right).
\end{eqnarray}
The potential Eq.~(\ref{pot}) is custodial $SU(2)$ symmetric if
\begin{eqnarray}
2\lambda_3&=&\lambda_2,\nn
2 \lambda_6 &=&2  \lambda_7 = \lambda_{11},\nn
\lambda_{9} &=& \lambda_{10}.
\end{eqnarray}

The Higgs vacuum expectation value causes a tree level mass splitting between the octet scalars. It is convenient to decompose the neutral complex octet scalars into two real scalars,
\begin{equation}
S^{A0}={S^{A0}_R+i S^{A0}_I \over {\sqrt 2}}.
\end{equation}
Then the tree level mass spectrum is,
\begin{eqnarray}
m^2_{S^{\pm}}&=&m_S^2+\lambda_1{ v^2 \over 4} ,\nn
m^2_{S_R^0}&=&m_S^2+\left(\lambda_1+\lambda_2+2\lambda_3\right){v^2 \over 4} ,\nn
m^2_{S_I^0}&=&m_S^2+\left(\lambda_1+\lambda_2-2\lambda_3\right){v^2 \over 4}.
\label{smass}
\end{eqnarray}

The Yukawa couplings of the octet scalars to the quarks are given in Eq.~(\ref{yuk8}). In the quark mass eigenstate basis these Yukawa couplings are
\begin{eqnarray}
L&=&-\sqrt{2}\eta_U \bar u_R^i {m_U^i \over v}T^{A}u_L^i  S^{A 0}+{\rm h.c.}\nn
&&+\sqrt{2}\eta_U \bar u_R^i {m_U^i \over v}T^{A}V_{ij}d_L^j S^{A +} +{\rm h.c.}\nn
&&-\sqrt{2}\eta_D \bar d_R^i {m_D^i \over v}T^{A}d_L^i S^{A 0 \dagger}+{\rm h.c.}\nn
&&-\sqrt{2}\eta_D \bar d_R^i {m_D^i \over v}V^{\dagger }_{ij}T^{A}u_L^j S^{A - } +{\rm h.c.}, 
\end{eqnarray}
where $V$ is the Cabibbo-Kobayashi-Maskawa matrix and $T^A$ are the eight color generators. Lower case Roman letters are used to denote mass eigenstate fields so, for example,  $u^1$ is the up quark field and $m_U^1$ is its mass. 

Which color octet state is the lightest depends on the parameters in the scalar potential. If a neutral state is the lightest, then the heavier charged ones can decay to it by emitting a charged $W$ boson. Whether the $W$ is real or virtual depends on the magnitude of the mass splitting. Similarly if the charged states are the lightest the two neutral states might be able to decay to them via emission of a charged $W$ boson. The color octet scalars can also decay to quarks. The charged ones decay almost exclusively to $t \bar b$ and the two neutral color octet scalars decay almost exclusively to $\bar t t$. 
\begin{eqnarray}
\Gamma(S^+ \rightarrow {t \bar b})&=&{|\eta_U|^2 \over 16 \pi m_S^3} \left({m_t \over v}\right)^2 |{V_{tb}}|^2\left(m_S^2-m_t^2\right)^2, \nn
\Gamma(S^0_{R} \rightarrow {t \bar t})&=&{m_S \over 16 \pi } \left({m_t \over v}\right)^2 \Biggl[\left|\text{Re\,}\eta_U\right|^2 \left(1-\frac{4m_t^2}{m_S^2}\right)^{3/2}\nn
&&+ \left|\text{Im\,}\eta_U\right|^2 \left(1-\frac{4m_t^2}{m_S^2}\right)^{1/2}\Biggr],\nn
\Gamma(S^0_{I} \rightarrow {t \bar t})&=&{m_S \over 16 \pi } \left({m_t \over v}\right)^2 \Biggl[\left|\text{Re\,}\eta_U\right|^2 \left(1-\frac{4m_t^2}{m_S^2}\right)^{1/2}\nn
&&+ \left|\text{Im\,}\eta_U\right|^2 \left(1-\frac{4m_t^2}{m_S^2}\right)^{3/2}\Biggr].
\end{eqnarray}

\section{Electroweak and Higgs Physics} 
\label{sec:matching}

For weak scale physics, the colored scalars can be integrated out and their effects are encoded in the coefficients of non-renormalizable operators that are $SU(3)\times SU(2)\times U(1)$ invariant. We are most interested in the non-renormalizable operators that, when added to those in the minimal standard model, change the rates for the processes $gg \rightarrow h$, $h \rightarrow \gamma \gamma$ and $h \rightarrow Z \gamma$. These processes are accessible at the LHC, and since in the standard model they arise
from one-loop matrix elements, they are sensitive to beyond the standard model physics. The total width for Higgs decay is approximately independent of the contributions of the higher dimension operators we are considering since for a light Higgs the total width is dominated by the channels, $h \rightarrow b \bar b$, $h \rightarrow c \bar c$,  $h \rightarrow \tau \bar \tau$, $h \rightarrow Z Z^*$ , $h \rightarrow W W^*$ which arise in the standard model from tree level matrix elements. For a heavier Higgs the decay rate is dominated by $h \rightarrow Z Z$ and $h \rightarrow W W$ which also occur at tree level and hence are insensitive to the effects of virtual color octet scalars.\footnote{Throughout this paper we assume $v^2/m_S^2$ is significantly less than unity. The center of mass energy of the LHC is $s=14$~TeV, which is larger than $m_S$. Nevertheless, one can still integrate out $S$ to compute its effect on Higgs production, because $gg\to h$ is governed by the parton 
center of mass energy $\hat s=m_h^2$.}

Assuming  CP conservation the correction to the Lagrange density from the dimension six operators that contribute to $\sigma(gg \rightarrow h)$, $\Gamma(h \rightarrow \gamma \gamma)$ and $\Gamma(h \rightarrow Z \gamma)$ are
\begin{eqnarray}
\label{lagrange}
\delta{\cal L}&=&-{c_G g_3^2\over 2 \Lambda^2} H^{\dagger} H G^A_{\mu \nu} G^{A \mu \nu}-{c_W g_2^2\over 2 \Lambda^2} H^{\dagger} H W^a_{\mu \nu} W^{a \mu \nu}\nn
&&-{c_B g_1^2\over 2 \Lambda^2} H^{\dagger} H B_{\mu \nu} B^{ \mu \nu} \nonumber \\
&&- {c_{WB} g_1g_2\over 2 \Lambda^2} H^{\dagger}\sigma^a H B_{\mu \nu} W^{a \mu \nu},
\end{eqnarray}
where $g_1$, $g_2$ and $g_3$ are the weak hypercharge, $SU(2)$ and strong $SU(3)$ gauge couplings, $B^{\mu \nu}$ is the field strength tensor for the hypercharge gauge group, $W^{a \mu \nu}$ is the field strength tensor for the weak $SU(2)$ gauge group and $G^{A \mu \nu}$ is the strong interaction field strength tensor. The first three terms  in Eq.~(\ref{lagrange}) are not constrained by precision electroweak physics. If we replace the Higgs field by its vacuum expectation value then they just give a redefinition of the gauge couplings. However the last term contributes to the precision electroweak variable $S$~\cite{Han}.

Integrating out the color octet scalars at one loop we find that,
\begin{eqnarray}
&&\frac{c_G}{\Lambda^2}={3 \over 2}\frac{c_W}{\Lambda^2}={3 \over 2}\frac{c_B}{\Lambda^2}=-{(2\lambda_1+\lambda_2)  \over 64 \pi^2 m_S^2},\nn
&&\frac{c_{WB}}{\Lambda^2}=-{\lambda_2  \over 48 \pi^2 m_S^2}\nn
&&\frac{c_{\gamma\gamma}}{\Lambda^2}\equiv \frac{c_W+c_B-c_{WB}}{\Lambda^2}=-{\lambda_1  \over 24 \pi^2 m_S^2}.
\end{eqnarray}
The coefficient $c_{WB}$ is constrained by the experimental results on the precision electroweak parameter $S$. Explicitly,
\begin{equation}
c_{WB}= -{1 \over 8\pi}{\Lambda^2 \over v^2} S.
\end{equation}
The current experimental limit is $S=-0.13 \pm 0.10$. The constraint on $S$ does not restrict the value of $c_G$ to be so small that it has a negligible impact on Higgs production at the LHC. For example, consider the case $m_S=0.75~{\rm TeV}$, $\lambda_1=4$ and $\lambda_2=1$. Then we have that $S\simeq 0.006$, $c_G=-0.03~ (\Lambda/{\rm 1~TeV})^2$ and $c_{\gamma\gamma}=-0.03~ (\Lambda/{\rm 1~TeV})^2$. A value of $c_G$ this large causes almost a factor of two increase in the Higgs production rate at the LHC~\cite{Manohar}, while decreasing the $H \to \gamma\gamma$ branching ratio by about 10\%. At present, the standard model gluon fusion Higgs production cross section has been calculated to NNLO order and a soft gluon resummation has been done \cite{nlo,nnlo,resum}. The theoretical uncertainty from higher order QCD corrections is estimated to be around $10\%$. There are also additional sources of uncertainly in the theoretical prediction for the cross section, for example, from the uncertainty in the parton distributions and the top quark mass. An enhancement by a factor of two is well outside the uncertainties in the standard model Higgs production rate.

The Higgs vacuum expectation value causes a tree level splitting between the charged and neutral $S$ scalars. This gives a contribution to the rho parameter,
\begin{equation}
\rho =\frac{M^2_{W^+}-M^2_{W^3}}{M^2_{W}}.
\end{equation}

The change in the rho parameter induced by the color octet scalars is,
\begin{eqnarray}
\Delta \rho &=& \frac{\alpha \dim}{16 \pi \sin^2\theta_W M^2_{W}}\nn
&&\times \left[
\frac12f(m_+,m_R)+\frac12 f(m_+,m_I)-\frac12f(m_R,m_I) \right]\nn
\label{rho}
\end{eqnarray}
where $m_+$ is the charged scalar mass, and $m_{R,I}$ are the neutral scalar masses and ${\rm dim}=8$ is the dimension of the color representation of the scalars. The function $f$ is,
\begin{eqnarray}
f(m_1,m_2) &=& m_1^2 + m_2^2 - \frac{2m_1^2m_2^2}{m_1^2-m_2^2} \ln \frac{m_1^2}{m_2^2}.
\end{eqnarray}
For $m_1 \simeq m_2$
\begin{eqnarray}
f(m_1,m_2) &\simeq& \frac{\left(m_1^2-m_2^2\right)^2}{3 m_1^2},
\end{eqnarray}
so that Eq.~(\ref{rho}) is approximately
\begin{eqnarray}
\Delta \rho &\simeq& \frac{\alpha}{6 \pi \sin^2\theta_W M^2_{W} m^2_{+,R,I}}
\left(m_+^2-m_R^2\right)\left(m_+^2-m_I^2\right)\nn
\end{eqnarray}
where $m^2_{+,R,I}$ is a typical scalar mass, and we have used $\dim=8$. Using the masses in Eq.~(\ref{smass}) gives
\begin{eqnarray}
\Delta \rho &\simeq& \frac{\alpha v^4}{96 \pi \sin^2\theta_W M^2_{W} m^2_S}
\left(\lambda_2^2-4\lambda_3^2\right)\nn
&=& \frac{\sin^2\theta_W M^2_{W}}{96 \alpha \pi^3   m^2_S}
\left(\lambda_2^2-4\lambda_3^2\right).
\end{eqnarray}

Note that the scalar contribution to the rho parameter is related to their contribution to the precision electroweak parameter $T$ by $\Delta \rho = \alpha \Delta T$. Suppose $m_S=0.75\, {\rm TeV}$, $\lambda_2=1$ and $\lambda_3=1$. Then the color octet scalars contribute $\Delta T=-0.04$ which is consistent with the experimental constraint, $T=-0.17 \pm 0.12$.

\section{Flavor Physics} 
\label{sec:flavor}

Virtual charged octet scalar exchange gives rise to new contributions to rare weak processes. For example, one loop diagrams give an additional contribution to the effective Hamiltonian for $|\Delta S|=2$ ~$K^0-\bar K^0$ mixing. This system has provided important constraints on extensions of the scalar sector that contain several (color singlet) Higgs doublets~\cite{Abbott}. Writing the new contribution to the effective Hamiltonian in the form,
\begin{equation}
\delta{\cal H}^{(|\Delta S|=2)}=C[V_{td}^*V_{ts}]^2(\bar d_L \gamma^{\nu} s_L)(\bar d_L \gamma^{\nu} s_L)+{\rm h.c.}, 
\end{equation}
we find that at one loop (neglecting, for simplicity, terms suppressed by $m_W^2/m_S^2$ and $m_t^2/m_S^2$) the Wilson coefficient $C$ is
\begin{eqnarray}
\label{mix}
&&C=|\eta_U|^2{M_W^2m_t^4 \over 24 \pi^2 v^4 (m_t^2-m_W^2)m_S^2}\left[1-{1 \over 4 }\left( {m_t^2 \over m_W^2}\right) \right. \nonumber \\
&&+ {1 \over 4(m_t^2-m_W^2)}\left(4M_W^2 {\rm ln}\left({M_W^2 \over m_t^2}\right) \right. \nonumber \\
&&+\left. \left.{ m_t^2 \over M_W^2}(m_t^2-2m_W^2){\rm ln}\left({m_S^2 \over m_t^2}\right)+m_W^2{\rm ln}\left({m_S^2 \over M_W^2}\right) \right) \right] \nonumber \\
&&+|\eta_U|^4 \frac{m_t^4}{16\pi^2 v^4 m_S^2} \frac{11}{18}.
\end{eqnarray}
Here we used the Fierz identities 
\begin{equation}
(\bar d_L \gamma^{\nu}T^A s_L)(\bar d_LT^A \gamma^{\nu} s_L)={1 \over 3}(\bar d_L \gamma^{\nu} s_L)(\bar d_L \gamma^{\nu} s_L),
\end{equation}
and
\begin{equation}
(\bar d_L \gamma^{\nu}T^A T^B s_L)(\bar d_LT^B T^A \gamma^{\nu} s_L)=\frac{11}{18}(\bar d_L \gamma^{\nu} s_L)(\bar d_L \gamma^{\nu} s_L).
\end{equation}
It is understood that in Eq.~(\ref{mix}) the Wilson coefficient is evaluated at a subtraction point of order $M_W$. We don't include renormalization group running between $m_S$ and $M_W$ but integrate out the color octet scalars, the top quark and the $W$-bosons at once. Since we are considering color octet scalars that are heavier than the Higgs doublet vacuum expectation value $v$, and $\eta_U$  is a free parameter, it is evident that  Eq.~(\ref{mix}) can give an acceptably small contribution to $K^0-\bar K^0$ mixing.

If $\eta_D$, is not small then the effects of virtual charged octet can be quite important for weak radiative B-meson decay. For small $\eta_U$, the effective Hamiltonian gets the additional contribution
\begin{equation}
\label{29}
\delta{\cal H}^{({\gamma})}=C_{\gamma} [V_{ts}^*V_{tb}]em_b {\bar s}_L \sigma_{\mu \nu}F^{\mu \nu} b_R+{\rm h.c.}~,
\end{equation}
with Wilson coefficient
\begin{equation}
\label{30}
C_{\gamma}=\eta_U^*\eta_D^*\left(m_t^2 \over v^2\right){1 \over 18 \pi^2m_S^2}f_{\gamma}(m_t^2/m_S^2),
\end{equation}
where
\begin{equation}
\label{31}
f_{\gamma}(x)={1 \over 4}\left({1+2x{\rm ln}x-x^2 \over (1-x)^3}\right)+\left({-1-{\rm ln}x+x \over (1-x)^2}\right).
\end{equation}

$\eta_U$ and $\eta_D$ also enter the effective Hamiltonian for nonleptonic $B$ decay via the gluon magnetic moment operator
\begin{equation}
\delta{\cal H}^{(g)}=C_{g} [V_{ts}^*V_{tb}]g_3m_b {\bar s}_L \sigma_{\mu \nu}G^{\mu \nu} b_R+{\rm h.c.},
\label{32}
\end{equation}
with Wilson coefficient
\begin{equation}
C_{g}=-\eta_U^*\eta_D^*\left(m_t^2 \over v^2\right){1 \over 24 \pi^2m_S^2}f_{g}(m_t^2/m_S^2),
\label{33}
\end{equation}
where
\begin{equation}
f_{g}(x)=\left({1+2x{\rm ln}x-x^2 \over (1-x)^3}\right)+{1 \over 4}\left({-1-{\rm ln}x+x \over (1-x)^2}\right).
\end{equation}

Since the charge $-1/3$ quarks all have masses much smaller than $v$, the value of $|\eta_D|$ can be much larger than unity without the color octet scalars being strongly coupled to the quarks. The inclusive $B \to X_s \gamma$ rate agrees with the standard model prediction to within the experimental and theoretical errors (about 10\%). Using Eqs.~(\ref{29}), (\ref{30}) and (\ref{31}) gives that the magnitude of the color octet scalar contribution to $C_{\gamma}$ divided by the standard model value is,
\begin{equation}
|C_{\gamma}^{({\rm octet})}/C_{\gamma}^{({\rm s.m.})}|=0.04\,|\eta_U \eta_D|f_{\gamma}(m_t^2/m_S^2)\left(1{\rm TeV}\over m_S\right)^2.
\end{equation}
Note that for $m_S \sim 1\,{\rm TeV}$ $f_{\gamma}\sim 3$. Since $|\eta_D|$ can be large even color octet scalars with masses well above a ${\rm TeV}$, and hence unobservable at the LHC,  can have a significant impact on weak radiative $B$ meson decay. The phases of $\eta_{U,D}$ can contribute to $CP$ violation in weak radiative $B$ meson decay. 

\section{Neutron Electric Dipole Moment}
\label{sec:dipole}

$CP$ violation in the couplings of the octet scalars contributes to the neutron electric dipole moment. The dominant contribution is through the color electric dipole moment of the $b$-quark~\cite{boyd}, which induces Weinberg's $CP$-violating three-gluon operator~\cite{weinberg} when  the $b$-quark is integrated out. 

The $b$-quark color electric dipole is given by computing the $b \to b g$ amplitude, rather than the the $b \to s g$ amplitude computed in Eqs.~(\ref{32}), so the result is obtained by the replacement ${\bar s}_L \to {\bar b}_L$ and $V_{ts}^* \to V_{tb}^*$ in Eq.~(\ref{32}), with Eq.~(\ref{33}) remaining unchanged. Using Eq.~(\ref{33}) at the scale $M_W$ running down to low energies~\cite{braaten,boyd} and computing the neutron matrix element of the three-gluon operator at low-energies using naive dimensional analysis~\cite{nda} gives us an estimate for the neutron electric dipole moment.

There are two versions of naive dimensional analysis which have been used~\cite{gunion}: (a) $g G_{\mu \nu}$ is counted as order $\Lambda_\chi^2$, or (b) $G_{\mu\nu}$ is counted as order $f_\pi \Lambda_\chi$. The two methods give the same result if one assumes that $\alpha_s(\mu\sim 1\, \text{GeV}) \sim 4 \pi$, otherwise, method (b) gives an estimate for the matrix element that is $[\alpha_s(\mu)/(4\pi)]^{3/2}$ times that of method (a). The renormalization group scaling of the operator to low energies also depends on $\alpha_s(\mu)$. Assuming $\alpha_s(\mu\sim 1\, \text{GeV}) \sim 1$, method (b) gives the estimate
\begin{equation}
d_n \sim 4\,\text{Im\,}\left[\eta_U^*\eta_D^* \right] 
   f_g(m_t^2/m_S^2) \left(1{\rm TeV}\over m_S\right)^210^{-26}\  e\text{-cm},
\end{equation}
and assuming $\alpha_s(\mu\sim 1\, \text{GeV}) \sim 4 \pi$, method (a) gives the estimate
\begin{equation}
d_n \sim \text{Im\,}\left[\eta_U^*\eta_D^* \right] 
 f_g(m_t^2/m_S^2) \left(1{\rm TeV}\over m_S\right)^210^{-26}\ e\text{-cm}.\nn
\end{equation}
For $m_S \sim 1\, \text{TeV}$, $f_g \sim 1.5$.
Note that the ratio of these two estimates is not $1/(4\pi)^{3/2}$ because of the anomalous dimension of Weinberg's $CP$-violating three-gluon operator. Even for color octet scalars with masses larger than a ${\rm TeV}$, the electron dipole moment of the neutron can be close to the experimental limit, $d_n<6.3 \times 10^{-26}\ e\text{-cm}$.

\section{Color Octet Production}
\label{sec:production}

In addition to changing the standard model values of physical quantities through their virtual effects, the color octet scalars can be produced directly at the LHC. Their production is dominated by gluon fusion and
the scalar production cross-section for either the charged or (two) neutral colored scalars is:
\begin{eqnarray}
&&\frac{{\rm d}\sigma}{{\rm d}t} = \frac{\pi \alpha_s^2}{s^2} {2 C_R d_R\over d_A^2} \left[ \left(C_R - \frac {C_A}4\right) +  \frac {C_A}4\frac{\left(u-t\right)^2}{s^2}\right] \nn
&&\times \left[1 + \frac{2m_S^2 t}{(m_S^2-t)^2}+\frac{2m_S^2 u}{(m_S^2-u)^2}
+\frac{4m_S^4}{(m_S^2-t)(m_S^2-u)}\right].\nn
\label{prod}
\end{eqnarray}
Here $C_{R,A}$ and $d_{R,A}$ are the quadratic Casimir and the dimension of the color representation for the final state scalars and for the adjoint representation,  respectively. For octet scalars, $C_R=C_A=3$, $d_R=d_A=8$. For real scalars, there is an additional factor of $1/2$. Eq.~(\ref{prod}) agrees with the squark production cross-section~\cite{Dawson} for $C_R=4/3$, $d_R=3$.

The integrated cross-section is
\begin{eqnarray}
\sigma &=& \frac{\pi\alpha_s^2}{s^2}{2 C_R d_R\over d_A^2} \Biggl[ \frac 1 6 \beta \left( 6 C_R (4m_S^2+s)+C_A(10m_S^2-s)\right) \nn
&&-\frac {4m_S^2} s \left(C_A m_S^2+C_R (s-2m_S^2)\right) \ln \frac{1+\beta}{1-\beta} \Biggr],
\end{eqnarray}
where
\begin{eqnarray}
\beta &=& \sqrt{1-\frac{4m_S^2}{s}},
\end{eqnarray}
is the velocity of the scalars in the center of mass frame.\footnote{
As $\beta \to 0$, the cross-section approaches
\begin{eqnarray}
\frac{\pi\alpha_s^2}{16 m^2}{2 C_R d_R\over d_A^2}(4C_R-C_A) \beta,
\end{eqnarray}
which is positive, since $C_R d_R (4 C_R - C_A)$ is always positive for non-singlet representations of any gauge group.}

\begin{figure}
\begin{center}
\includegraphics[width=9.0cm]{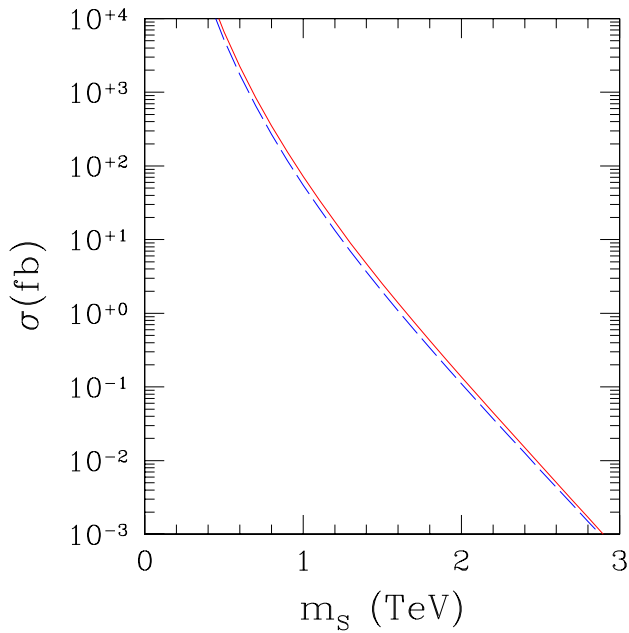}
\end{center}
\vspace{-0.75cm}
\caption{Production rate for color octet scalars at the LHC via gluon fusion as a function of the scalar mass. The rate shown is that for charged scalars. One needs to use half the above rate for each of the two neutral scalars. The rates were computed using the CTEQ5 distributions at LO (dashed blue) and NLO (solid red).
\label{fig:rate}}
\end{figure}

In Fig.~\ref{fig:rate}, we have plotted the charged scalar production rate as a function of the scalar mass. The hadronic cross-sections were computed using the CTEQ5 parton distributions~\cite{cteq}, and $\alpha_s$ was evaluated at $\mu=2m_S$. The neutral scalars $S_R^0$ and $S^0_I$ are real, and so each of them has a production rate half that of a charged scalar with the same mass.

Observability of the color octet scalars depends on their decays. Consider, for example, the situation where the charged color octet scalars are heavier than the neutral ones and can decay to the neutral ones via W boson emission. Then the decay chains, $S^{+}\rightarrow W^{+}S^0 \rightarrow W^{+}t \bar t  \rightarrow W^{+}W^+W^- b \bar b $, $S^{+}\rightarrow t \bar b \rightarrow W^{+} \bar b b$, and $S^0 \to t \bar t \to W^+ W^- b \bar b$ occur.

\section{Unification}
\label{sec:unify}

Simple unification scenarios for the standard model do not give the precise experimental value of $\sin^2\theta_W$, and have a unification scale that is too small for an acceptable proton decay rate. In standard $SU(5)$ unification of the minimal standard model,  the one loop running gives $\sin^2\theta_W = 0.207$ and $M_G = 7 \times 10^{14}$~GeV using $\alpha_s(M_Z)=0.1187$. Since $\sin^2\theta_W$ is known more accurately than $\alpha_s(M_Z)$, it is better to use $\sin^2\theta_W = 0.2312$ as the input, and then predict  $\alpha_s(M_Z)=0.07$ and $M_G = 1 \times 10^{13}$~GeV. Adding the  $\rep81{1/2}$ octet scalar changes the above values to $\alpha_s(M_Z)=0.06$ and $M_G = 8 \times 10^{13}$~GeV. This gives a somewhat better value for $M_G$, but $\alpha_s(M_Z)$ is about the same as in the standard model.

Unification can be achieved by adding other scalars that do not couple to the quarks and leptons. Consider, for example, adding to the minimal standard model the $\rep81{1/2}$ octet scalars and $n_T$ real $\rep13{0}$ scalars. Then for $n_T=3$, $\alpha_s(M_Z)=0.122$ and $M_G= 9 \times 10^{15}{\rm GeV}$.

The additional $SU(2)$ triplets add the following terms to the most general renormalizable scalar potential,
\begin{eqnarray}
\Delta V&=&m_{T_j}^2T_j^a T_j^a + \lambda_{jklm}T^a_jT^a_kT^b_lT^b_m\nn
&+&\xi \epsilon_{ijk}\epsilon^{abc}T^a_iT^b_jT^c_k +  {\mu_{1j}}(S^{A\dagger} \sigma^a S^A)T^a_j \nonumber \\
&+&g_{jk}H^{\dagger} HT_j^aT_k^a +{\mu_{2j}}(H^{\dagger} \sigma^a H)T^a_j.
\end{eqnarray} 
where $j=1-3$ labels the different triplets.
We have suppressed the $SU(2)$ indices on the Higgs doublet and on the color octet scalars. Brackets are sometimes used to indicate how the suppressed $SU(2)$ indices are contracted. We assume that there is an approximate $T \rightarrow -T$ symmetry that is broken by the very small dimensionful parameters $\mu_1$ and $\mu_2$ as well as the dimensionful parameter $\xi$. For simplicity, we take the triplet scalars to be very heavy so that their production at the LHC is negligible.

Since the parameters that break the $T \rightarrow -T$ symmetry are very small, the vacuum expectation value of the neutral component of these fields can be neglected. Nonetheless, the $T\rightarrow-T$ symmetry breaking is needed for these scalars to decay. 

The above scenario gives acceptable coupling constant unification and proton decay, with only a modest extension of the low-energy scalar sector of the standard model.

\section{Conclusions}
\label{sec:conclude}
The scalar sector of the standard model may be more complicated than just a single Higgs doublet. We argued that natural suppression of flavor changing neutral currents implies that only scalars with $SU(3) \times SU(2) \times U(1)$ quantum numbers of the Higgs doublet or $\rep82{1/2}$ will couple to the quarks. Motivated by this we considered the phenomenology of a model where a $\rep82{1/2}$ scalar is added to the minimal standard model. Even though this model has no flavor changing neutral currents at tree level, it does give corrections from loop graphs, that may be important for precision flavor physics. In particular there are new sources of $CP$ violation which give rise to a contribution to the electric dipole monent of the neutron that can be near the present experimental limit.

For a range of parameters, the modification of the standard model we propose can have a significant impact (factors of two) on the Higgs boson production rate at the LHC. The spectrum of the color octet scalars depend on parameters in the Higgs potential, and it influences the decay channels available to these scalars. For a range of parameters, the effects of these colored scalars would first be seen indirectly at the LHC via their influence on the Higgs production rate, rather than via direct colored scalar production.
Finally we showed that unification of the gauge couplings occurs in our model if one also adds three real scalars with $SU(3) \times SU(2) \times U(1)$  quantum numbers $({\bf 1},{\bf 2})_0$.

We thank M. Ramsey-Musolf for useful discussions on the electric dipole moment of the neutron. We also thank C.~Anastasiou, M. Graesser, D. Hitlin, G. Isidori, and  Z. Ligeti for comments.

\begin{appendix}
\section{Naturalness}
\label{app:natural}

MFV leads to neutral current flavor conservation. The converse is not necessarily true. Neutral current flavor conservation implies that all the Yukawa matrices are diagonal in the same basis. MFV requires that they are all proportional (to $g^U$ or $g^D$), which is a stronger constraint. We have argued that the additional requirement of \emph{naturalness}  implies the stronger constraint that all the Yukawa couplings are proportional to $g^U$ or $g^D$, and hence implies MFV. 

What is natural depends on assumptions about the underlying theory and how parameters are generated. We have assumed that symmetries are important, and that the underlying theory has a $SU(3)_U \times SU(3)_D \times SU(3)_Q$ flavor symmetry that is broken, and this breaking generates Yukawa matrices. There are other scenarios which would lead to different conclusions. For definiteness, consider two Yukawa matrices $A$ and $B$ which transform like $g_U$. They can be written as
\begin{eqnarray}
A = U_A \Lambda_A V_A,\qquad
B = U_B \Lambda_B V_B
\end{eqnarray}
where $U_{A,B}$ and $V_{A,B}$ are unitary, and $\Lambda_{A,B}$ are real, non-negative and diagonal. It is possible that the underlying theory generates $\Lambda_{A,B}$ by some mechanism, and $U_{A,B}, V_{A,B}$ by a different mechanism which naturally gives $U_A=U_B$ and $V_A=V_B$. Then one would automatically have no tree-level flavor changing currents, but the MFV constraint would not be satisfied since $\Lambda_A$ and $\Lambda_B$ need not be proportional.

Note that a simple algebraic relation such as $\left[A,B\right]=0$ is not equivalent to the requirement that $A$ and $B$ are simultaneously diagonalizable, since $A$ and $B$ are not hermitian. Simple counterexamples for two flavors are
\begin{eqnarray}
A = \left( \begin{array}{cc} 1 & 0 \\ 1 & 0 \end{array} \right),\qquad
B = \left( \begin{array}{cc} 1 & 0 \\ 0 & 1 \end{array} \right)
\end{eqnarray}
which commute but are not simultaneously diagonalizable, and
\begin{eqnarray}
A = \left( \begin{array}{cc} \cos \theta & 0 \\ -\sin \theta & 0 \end{array} \right),\qquad
B = \left( \begin{array}{cc} 0 & \sin \theta \\ 0 & \cos \theta \end{array} \right)
\end{eqnarray}
which are simultaneously diagonalizable but do not commute.

\end{appendix}

\end{document}